\documentclass[12pt]{article}
\pdfoutput=1
\usepackage{epsfig}
\usepackage{amsfonts}
\usepackage{amssymb}
\usepackage{xcolor}
\usepackage{amsmath}
\usepackage{upgreek}
\usepackage{graphicx}

\begin{document}
\begin{center}

{\bf \Large Generalising holographic fishchain}\vspace{1.0cm}

{\bf \large R.M. Iakhibbaev$^{1}$, D.M. Tolkachev$^{1,2}$ \\[0.3cm]} \vspace{0.5cm}

{\it $^1$Bogoliubov Laboratory of Theoretical Physics, Joint Institute for Nuclear Research, Dubna

$^2$ Stepanov Institute of Physics, Minsk, Belarus}
\vspace{0.5cm}

\abstract{In this paper we present an attempt to generalise the integrable Gromov-Sever models, the so-called fishchain models, which are dual to biscalar fishnets. We show that in any dimension they can be derived at least for some integer deformation parameter of the fishnet lattice. We focus in particular on the study of fishchain models in AdS$_7$ that are dual to the six-dimensional fishnet models}
\end{center}

\section*{Introduction}
Due to developments in string theory and quantum field theory, the idea of holography has become widely implemented in modern physics. AdS/CFT duality in its various forms represents the brightest example of the correspondence between the weak coupling constant regime in quantum field theory and the strong regime in gravitation. At the same time, direct proof of the connection between the theories in bulk and at the boundary has not yet been found  either, and the direct connection between the theories is still unclear.  Within integrable models, a direct derivation of duality from first principles without supersymmetry --- the duality between fishnet and fishchain models  --- was presented not so long ago \cite{GromovSever, GromovAdS,GromovGamma}.

The biscalar fishnet theory is a specific $\mathcal{N}=4$ SYM limit preserving integrability and conformality in the planar limit \cite{Caetano, integrabfishnet}. This theory is obtained by strong $\gamma$-deformation of $\mathcal{N}=4$ and a double scale limit where the twist parameters become  large and the Yang-Mills coupling constant $g_{YM}$ is assumed to be  small \cite{Grabner:2017pgm}. The Lagrangian of biscalar fishnet theory is given by \cite{integrabfishnet}
\begin{equation}
     \mathcal{L}=N_c ~ \text{Tr}\left(-\phi^\dag \square \phi-\varphi^\dag \square  \varphi +  (4\pi)^2 \xi^2 \phi^\dag \varphi^\dag \phi \varphi \right)
\end{equation}
where both scalar fields transform under the adjoint representation of $SU(N_c)$, $\xi^2$ is the coupling constant, $\square$ is the D'Alembert's operator. The Feynman diagrams in the planar limit of a model like this form a square lattice of integrals. This model is exactly solvable as the corresponding Heisenberg spin chain \cite{Zamolodchikov}, and hence one can calculate the whole spectrum of the anomalous dimension and obtain a correlation function in any limit for the coupling constant within the framework of the Bethe-Salpeter approach \cite{ExactCorr, Derkachov2019}. 

The conditions imposed on anomalous dimension, on the other hand, can be interpreted as physical ones, and analyzing them, we can build a dual theory for the biscalar-fishnet model, the so-called fishnet model, the classical Lagrangian of which can be represented as follows:
\begin{equation}
    L=\xi^2 \sum_{i=0}^J  \left(\frac{\dot{X}^2_i}{2} +\prod_{i=1}^J  (-X_i \cdot X_{i+1})^{1/J}\right)
\end{equation}
where $X_i(t)$ are the world-sheet coordinates of the $i$-scalar particle on the projective lightcone in $\mathbb{R}^{1,5}$. In the works of Gromov and Sever it was shown how exactly this model is quantized and how exactly the duality between the fishchain model in AdS$_5$ and the fishnet model as CFT$_4$ can be established \cite{GromovAdS}. It was also shown that it is possible to generalise the results to a more complex case of the fishnet model with an appropriate integrable holographic model including magnons and antimagnons\cite{GromovGamma}.  

At the same time, there exists a natural generalization of the biscalar fishnet model to arbitrary dimensions \cite{FishnetAnyDim}, and for this generalization one can also obtain an exact anomalous dimension and correlation functions. In the framework of the holographic principle, Of course, it would be interesting to show the duality between generalised fishnet theories and the fishchain models on arbitrary dimensions. 

Intriguing, among these models, there is is non-isotropic six-dimensional fishnet model of theory corresponding to the ladder limit of the  6d $\mathcal{N}=(1,1)$ SYM theory in the planar limit \cite{BorkAmp}. In this limit the theory is convergent and it is possible to find a closed expression for the on-shell amplitude and correlation function \cite{Bork:2015zaa,Broadhurst:2010ds}. Moreover, it is well-known that the 6d SYM (1,1) is an effective low-energy limit of string theory describing D5-brane dynamics, that is why obtaining a dual model for 6d fishnet theories seems a step toward a full dual description of higher-dimensional $\mathcal{N}=2$ SYM-like models. Throughout the paper we will focus on six-dimensional fishnet models.

The paper is organised as follows. In the first part, we give an introduction to the generalised biscalar fishnet model in arbitrary dimensions and outline its main features. Then, in the second section, we show  how to get a holographic fishchain models dual to a set of biscalar fishnet models.  In the third section we proceed to applications where we compute spectra and test the validity of dual models on some examples. In the last section, we give a direct proof of a fishnet/fishchain correspondence. 

\section{Biscalar fishnet model in arbitrary dimension}

In general, biscalar models in an arbitrary number of dimensions can be written in the form of the following density Lagrange function \cite{FishnetAnyDim,BorkAmp,KazakovLoom}:
\begin{equation} \label{fishnetLagr}
    \mathcal{L}=N_c \text{Tr}\left(\phi^\dag (-\square)^\omega \phi +\varphi^\dag (-\square)^{d/2-\omega} \varphi + \xi^2 (4\pi)^{d/2} \phi^\dag \varphi^\dag \phi \varphi \right)
\end{equation}
with an arbitrary lattice isotropy parameter $\omega \in (0,d/2)$ and space-time dimension $d$.  In the case of arbitrary parameters $(d,\omega)$, the model (\ref{fishnetLagr}) is non-local.  At the quantum level, the action (\ref{fishnetLagr}) is not complete and requires counter-terms that are double-trace operators:
\begin{equation}
    \mathcal{L}_{dt} \sim \alpha_i(\xi) \sum_i \text{tr} (O_2) \text{tr} (\Tilde{O}_2)
\end{equation}
where $O_2$ and $\Tilde{O}_2$ are monomials consisting of $(\phi,\varphi)$ and their conjugations. Obviously, by introducing such counter terms, the model becomes non-conformal, but by finding a UV fixed point of $\alpha_i(\xi)$, the conformal symmetry of the model can be restored \cite{FishnetAnyDim}. This model cannot be derived from any higher-dimensional theory, unlike the situation in four dimensional space-time, where the biscalar fishnet model appears as the double scaling limit of the $\gamma$-deformed $\cal{N}$=4 SYM as noted in the introduction.  However, such a $d$-dimensional model in planar limit is also integrable due to its equivalence with the integrable spin chain $SO(1,d+1)$ \cite{FishnetAnyDim}.The lattice structure of this biscalar fishnet model in the planar limit is regular but non-isotropic due to the deformation parameter $\omega$ in the Lagrangian, so this parameter can be called the non-isotropic parameter. 

\begin{figure}[h]
\begin{center}
\includegraphics[scale=0.15]{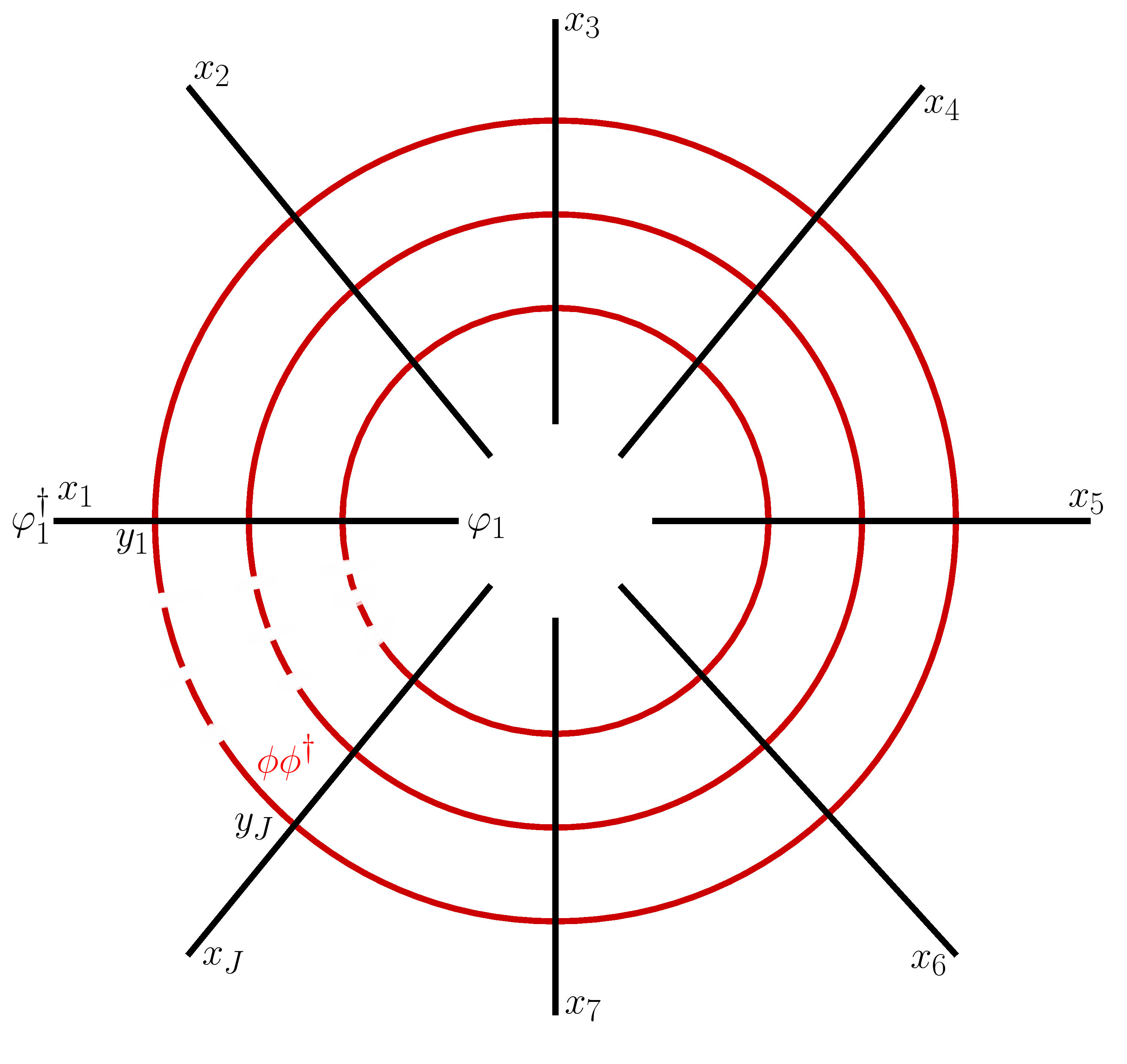}
\caption{Wheel-graph corresponding to the typical form of zero $U(1)$-charge CFT-wavefunction. Red lines corresponds to the neutral combination of $\phi \phi ^\dag$ }
\end{center}
\label{wheelgraph}
\end{figure}

The regular lattice structure allows the iterating element to separate. The graph-building operator in this model can be written by defining its integral kernel \cite{FishnetAnyDim}
 \begin{equation}
    \hat{B}= \prod_{i=1}^J \frac{ \; \xi^2/\pi^{d/2}}{|\vec{y}_{i}-\vec{y}_{i+1}|^{d - 2\omega} |\vec{x}_i-\vec{y}_i|^{2\omega}}
\end{equation}
\begin{figure}[h]
\begin{center}
\includegraphics[scale=0.2]{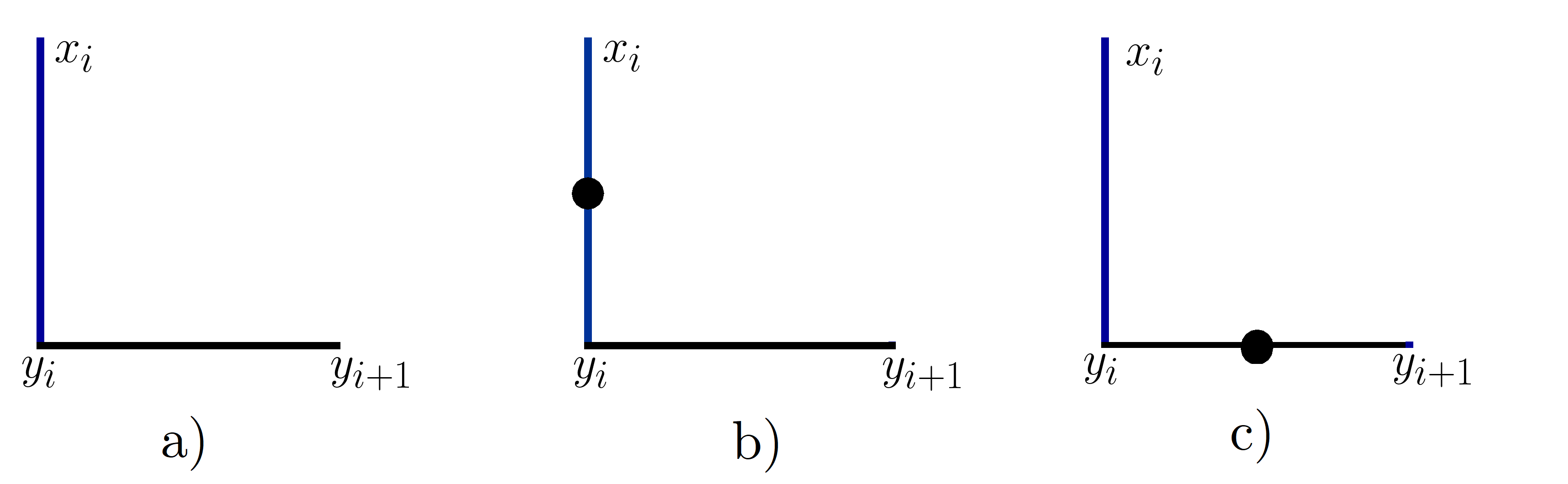}
\caption{Examples of fishnet graph-building operators for a) graph-building operator for isotropic fishnet, b) 6d $\omega=1$ fishnet model, c) 6d $\omega=2$ fishnet model. Dot denotes doubled propagator. Different colours of the lines correspond to different propagators of the appropriate fields }
\end{center}
\label{fishnets}
\end{figure}
This integral can be thought of as an operator that acts on the lattice, adding another layer to the Feynman integral. The eigenvector of this operator (vector on Hilbert space) is called the CFT wave function $\Psi$. This function in the general form can be defined as
\begin{equation}\label{cftwf}
    \Psi_\mathcal{O}=\langle \mathcal{O}_J(x_0) \; \text{tr}[\varphi^\dag(x_1) \ldots \varphi^\dag(x_J)] \rangle
\end{equation}
with $\mathcal{O}_J$ being a local operator, i.e it is just the $J+1$-point correlation function. This function can be written in the form having zero $U(1)$ charge:
\begin{equation}
    \mathcal{O}_J=\text{tr}[\partial^A \varphi (\phi^\dag \phi)^B]+\ldots
\end{equation}
with a neutral combination of $\phi$ and its conjugations (if the $U(1)$ charge of the single-trace operator is non-zero, such an operator is said to contain magnons). In the planar limit, such a correlation function contains Feynman diagram of a special kind \cite{ExactCorr}, the so-called wheel graph, example of which is depicted on Fig. \ref{wheelgraph}.

The explicit form of the wave function for a two-particle interaction case can be fixed by conformal symmetry that is why it depends on eigenvalues of dilatation operator $\Delta$. This reflects that all conformal generators and the graph-building operator commute. The eigenvalues of the graph-building operator represent the spectrum (in the strict sense it is a null-magnon spectrum) of the model which is defined by the following equation:
\begin{equation}
    \hat{B} \circ \Psi_{\Delta,S} = h^{-1}_{\Delta,S} \Psi_{\Delta,S}
\end{equation}
where $h$ denotes a spectrum.
The detailed method for obtaining the spectrum is presented in the appendix, here we give an explicit expression for the spectrum:
\begin{equation}\label{spectrumforall}
  h_{\Delta,S}= \frac{   \Gamma (\omega )^2}{\Gamma \left(\frac{d}{2}-\omega \right)^2 } \frac{ \Gamma \left( \frac{d}{4}+\frac{S}{2}-\frac{\Delta}{2} \right) \Gamma \left(\frac{d}{4}  -2 \omega  + \frac{S}{2} + \frac{\Delta}{2}\right)}{\Gamma \left(\frac{d}{4}+\frac{S}{2}+\frac{\Delta}{2}\right) \Gamma \left(\frac{d}{4}  +2 \omega  + \frac{S}{2} - \frac{\Delta}{2} \right)}
\end{equation}
Hereafter, we will concentrate on integer values of the deformation parameter and the spacetime dimension. For example, for $d=4$ and $\omega=1$ one can find the well-known result \cite{integrabfishnet}:
\begin{equation}
    h_{\Delta,S}^{-1}=\frac{1}{16} (-\Delta +S+2) (-\Delta +S+4) (\Delta +S-2) (\Delta +S)
\end{equation}
 and also one can find another spectrum with $d=6$ and $\omega=1$ \cite{BorkAmp}:
\begin{equation}\label{spec61}
    h_{\Delta,S}^{-1}=\frac{1}{16} (-\Delta +S+3) (-\Delta +S+5) (\Delta +S-1) (\Delta +S+1).
\end{equation}
So we can conclude, the physical spectra of these two models are the same under shift of the conformal spin.
For $d=6$ and $\omega=2$, the situation is different and spectrum has the following form:
\begin{multline}\label{spec6d2}
   h_{\Delta,S}^{-1}= \frac{1}{256}(\Delta -S-3) (\Delta -S-5) (\Delta -S-7) (\Delta -S-9) \\ (\Delta +S-5)  (\Delta +S-3) (\Delta +S-1) (\Delta +S+1)
\end{multline}
reminding a ''squared spectrum''. Within the dual description that will be given in the sections below, this spectrum is of a more comprehensible nature.

As it was said earlier each action of $\hat{B}$ on $\Psi$ adds another layer to the lattice, thus creating a geometric series that can be formally summarised:
\begin{equation} \label{geomprog}
1+\hat{B}+\hat{B}^2+\ldots=\frac{1}{1-\hat{B}}
\end{equation}

After all, such an expression allows the four-point correlation function to be written in a very compact form:
\begin{equation}
    \langle x_1,x_2| \frac{1}{1-\hat{B}} |   y_1,  y_2\rangle = \sum_S\int d\Delta  \frac{1}{1-h^{-1}_{\Delta,S}}   \langle x_1,x_2| \Psi_{\Delta,S} \rangle \langle \Psi_{\Delta,S} | y_1, y_2\rangle
\end{equation}
In fact, the equation for the poles can be represented in the form of equation:
\begin{equation}
    \left(1-\hat{B}\right)\circ \Psi_{\mathcal{O}}=0
    \label{physcond}
\end{equation}
The integrand in brackets can be transformed into a conformal block with an appropriate measure of integration (its form is not very important in this discussion but one can find it in, e.g., \cite{CFTharm}). Thus, the final result after integration by residues is a correlation function in the familiar form \cite{cftope}:
\begin{equation}
    G(u,v)= \sum_{S,\Delta} C_{\Delta,S} ~g_{\Delta,S}(u,v) 
\end{equation}
where $u,v$ are the corresponding conformal variables, $C_{\Delta,S}$ is the structure constant and $g_{\Delta,S}(u,v)$ is the conformal block which has a fixed form \cite{cftope}. Equation (\ref{physcond}) for the graph-building operator plays a key role because it selects ''physical states''. Indeed, it turns out that interpreting it as a physical condition, one can find a holographically dual picture. Let us explore the holographic side of the model.

\section{Fishchain from biscalar model}
 
As stated earlier, the condition imposed on the CFT wave functions can be interpreted as physical. So the zeros of the denominator in the operator expression (\ref{physcond}) can be thought of as stationarity conditions of the wave function. 
Acting on the equality (\ref{physcond}) with the D'Alamber operator $(-\square_i)^\omega$ and using the following identities
\begin{equation} \label{dif1}
    (-\square)^\omega D(x-y)=\delta^d(x-y),
\end{equation}
\begin{equation}\label{dif2}
    D(x-y)=\frac{\Gamma(d/2-\omega)}{4^\omega \pi^{d/2} \Gamma(\omega)} \frac{1}{\left((x-y)^{2}\right)^{\frac{d-2 \omega}{2}}},
\end{equation}
one can notice its similarity to the equation of the theory that has invariance with respect to time reparametrization:
\begin{equation} \label{hamiltclas}
      \hat{H} ~\Psi(\{x_i\}) =0, 
\end{equation}
where the Hamiltonian can be identified with
\begin{equation} \label{hampart}
    \hat{H}=\prod_i p_i^{2\omega} - \prod_i \frac{4^\omega \xi^2}{((x_i-x_{i+1})^2)^{\omega}} \frac{\Gamma(\omega)}{\Gamma(d/2-\omega)}
\end{equation}
because of $p_x = -i \partial_x$ (wherever not highlighted separately, the product is counted from 1 to $J$). Obviously, this Hamiltonian, by virtue of equation (\ref{hamiltclas}), has a zero spectrum. This symmetry allows us to formulate the model in a slightly different form. To see that, let us turn to Lagrangian, which corresponds to Hamiltonian (\ref{hampart}): 
\begin{equation}
    L=\frac{2 J \omega -1}{(2\omega)^{\frac{2J \omega}{2 J \omega -1}}} \left(\frac{1}{\gamma} \prod_i \dot{x}_i^{2 \omega}\right)^{\frac{1}{2 J \omega -1}}+\gamma \prod_i\frac{4^\omega \kappa \xi^2}{((x_i-x_{i+1})^2)^{\omega}}
\end{equation}
where the gauge parameter $\gamma$ transforms under time-reparametrisations $t \rightarrow f(t)$ as $\gamma \rightarrow \gamma/f'(t)$, and $\kappa$ is for the combination of gamma-functions. One can fix gauge $\gamma$ by extremising it:
\begin{equation}
    S=2 J\xi^{\frac{1}{\omega}} \kappa^{\frac{1}{2\omega}}\int dt     \left(\prod_i \frac{\dot{x}_i^{2\omega}}{((x_i-x_{i+1})^2)^{\omega}}\right)^{\frac{1}{2 J \omega}}
\end{equation}
Here and below we assume that $\omega$ is integer or half-integer. 

It is well known that conformal algebra on $d$ dimension is the same as the algebra of $SO(1,d+1)$ and the $d$ dimensional conformal invariant quantity is a $d+2$ scalar on $SO(1,d+1)$.
In this identification, the flat space coordinate $x^{\mu=-1,0....n}$ is mapped to the projective lightcone of $\mathbb{R}^{1,{d+1}}$.
$x^\mu=X^\mu_i/X^+_i$, $X^{+}_i=X^0_i+X^{-1}_i$, and lightcone constraint will be given by $X^2_i=0$. By doing this embedding, a Nambu-Goto-type action for a ''chewed'' or discretised string can be found:
\begin{equation}\label{ngtype}
    S=2 \xi^{\frac{1}{\omega}} \kappa^{\frac{1}{2\omega}} J \int dt   \left(\prod_i \frac{\dot{X}_i^{2 \omega}}{(-2X_i \cdot X_{i+1})^\omega}\right)^{\frac{1}{ 2 \omega J}}
\end{equation}
It is easy to see that this action in a particular case reproduces a four-dimensional result and corresponds to the t'Hooft holographic principle. At the same time, the form of the action itself is completely identical to that found by Gromov and Sever \cite{GromovSever}.
To proceed, one can find a form for such a Lagrangian in the Polyakov type discretised action \cite{GromovSever,openfish}: 
\begin{equation} \label{poltype}
    L=\sum_i \alpha_i\frac{P^2_i}{2} - \prod_k (- \alpha_i X_i \cdot X_{i+1})^{-\frac{1 }{J}}+\eta_i X_i^2
\end{equation}
where $\alpha_i$ and $\eta_i$ are the Lagrange multipliers. The first multipliers are responsible for global gauge transformation and the second one keeps the particle on the lightcone. To proceed, one should fix the gauge as these auxiliary field are transformed under the symmetries of the action. There are a few of them: 1) conformal symmetry, 2) time-reparametrization symmetry, 3) translation along the closed chain $X_i \rightarrow X_{i+1}$, 4) time-dependent rescaling due to uplifting $X_i \rightarrow g_i(t) X_i, ~ \alpha_i \rightarrow g^2_i \alpha_i, \eta_i \rightarrow g^{-2}_i \eta_i$. After extremizing the action by $\alpha_i$ one should choose  $\alpha_i=1$ to fix the gauge symmetry of the fishchain as follows:
\begin{equation}
    \mathcal{L}=2\prod_i (-X_i \cdot X_{i+1})^{1/J}=\dot{X}^2_k
\end{equation}
This expression can be thought of as a constraint on the dynamics of a discretised string: energy density along this string bit is zero and therefore it is associated with the first Virasoro constraint.  There is one more constraint which can be fixed as $\mathcal{L}=m^2$ to be a constant due to the time reparametrization $t \rightarrow f(t)$ and the corresponding rescaling $X_i(t) \rightarrow  X_i(t) {f'}^{-1/2}$, $\eta_i \rightarrow \eta_i f'^2$ symmetries.   

The action (\ref{ngtype}) has exactly the same form as the action for the six-dimensional fishchain theory, so it inherits all of its properties being defined in $D=d+2$ space-time.  From the Lagrangian form itself, one can directly see the form of the correlation function at the classical level and scaling dimension $\Delta \sim \xi^{1/\omega}$ for $\xi \gg 1$, which can be found from (\ref{spectrumforall}). This reminds the quasiclassical limit in quantum mechanics where expansion is performed by the inverse Planck constant. 

Anticipating the quantization of this model, it is convenient to introduce an invariant $SO(1,d+1)$-charge density  
\begin{equation}
    q^{MN}_i=2P^{\left[M \right.}_i \cdot X^{\left. N \right]}_i 
\end{equation}
and express the obtained quantities in accordance with this invariant. The global $SO(1,d+1)$ charge is defined as $Q^{MN}=\xi^{1/\omega} \sum_i q^{MN}_i$, while the square brackets denote the commutation of indices. As theory conformal at the classical level we can choose the scaling dimension $Q^{-1,0}=i\Delta$ and the Lorentz spin as $Q^{1,2}=S_1$, $Q^{3,4}=S_2$. 

One can reformulate equations of motion in terms of the charge density in the following form:
\begin{equation}
    \dot{q}^{MN}_i= \frac{\mathcal{L}}{2} (j^{MN}_{i+1}-j^{MN}_i)
\end{equation}
where the current density\footnote{Here one can draw an analogy with similar charges in sigma models \cite{SM}. This expression  can further support the connection between sigma models and fishchain models in the discussion of the continuum limit of biscalar fishnet models \cite{Bassosigma}}

\begin{equation}    j^{MN}_i=\frac{X^{\left[M \right.}_i \cdot X^{\left. N \right]}_{i+1}}{X_i \cdot X_{i+1}}
\end{equation}
There is also an interesting relation between charge density and current density:
$$
j_{i+1}q^2_i-q^2_i j_i=0
$$
Using these two last equations, one can find a zero-curvature relation and an appropriate Lax pairs to proof integrability of the classical fishchain  \cite{GromovSever}.

The constraints imposed on the system of string-bits can be rewritten in the following form:
\begin{equation}
C_X=X_i^2, ~ C_{P}=\frac{1}{2}(P_i^2-m^2), ~ C=X_i \cdot P_i
\end{equation}
with $m^2=\mathcal{L}$ fixed to be a constant in the appropriate gauge. The corresponding commutation algebra can be written with the Dirac brackets as the last two conditions are second class constraints. In terms of the charge densities, one can easily find a Hamiltonian in invariant form which is convenient in use and quantization \cite{GromovAdS}.

\section{Quantization of fishchain}

Quantization of such a theory is carried out in full compliance with the six-dimensional fishchain which was outlined in \cite{GromovAdS} considering all the first and second-class constraints imposed on the system of string-bits. It is possible to construct and specify canonical commutative relations in the same way as in \cite{GromovAdS,GromovGamma}  with a simple replacement $\xi \rightarrow \xi^{1/\omega}$:
\begin{equation}\label{crx1}
    [\hat{X}^M_m, \hat{P}^N_n]=\frac{i \delta_{mn}}{\xi^{1/\omega}} \left[\eta^{M N}-\frac{1}{m^2} \hat{P}^M_k \hat{P}^N_k\right]
\end{equation}
\begin{equation}\label{crx2}
    [\hat{P}^M_m, \hat{P}^N_n]=0
\end{equation}
\begin{equation}\label{crx3}
    [\hat{X}^M_m, \hat{X}^N_n]=\frac{i \delta_{mn}}{\xi^{1/\omega}} \frac{\left[\hat{P}_k^{M},\hat{X}_k^{N}\right]}{m^2}
\end{equation}
One may again notice a similarity between the coupling constant and the Planck constant: this is a consequence of our interpretation of the classical action where the coupling constant enters as a factor in the style of the t' Hooft holographic principle. The algebra seems cumbersome to carry out calculations. It is convenient to refine the operators as follows:
\begin{equation}
    \hat{X}^M_i=-\frac{1}{m^2 \xi^{1/\omega}} \hat{P}^{M}_i + \hat{Y}^M_i-\frac{1}{m^2} \left(\hat{P}_i\cdot \hat{Y}_i\right) \hat{P}^M_i
\end{equation}
after such a shift the constant in the denominator of (\ref{crx1}-\ref{crx3}) is eliminated and the commutation relations can be reduced to a more familiar form 
\begin{equation}
    [\hat{Y}^M_m, \partial_{Y,n}^N]=\frac{i}{\xi^{1/\omega}} \eta^{M N} \delta_{m n}, ~\; [\hat{Y}^M_m, \hat{Y}_{n}^N]=0
\end{equation}
where $Y_i=Z_i R$ is coordinate in $AdS_{D-1}$ space, which appears as a consequence of quantizing procedure (here $\hat{Y}$ acts just as the multiplication, thus $Y_i$ are usual coordinates of particles in the fishchain). $\partial_{Y,n}^N$ is the corresponding derivative. These operators act on the vectors of Hilbert space: vectors in coordinate representation is a dual CFT wave functions $\widehat{\Psi}(\{Y_i\})$. The norm of these dual vectors and direct proof of duality between $\widehat{\Psi}(\{Y_i\})$  and $\Psi (\{x_i\})$ will be given in the last section.

Emergent radius of $AdS_{D-1}$ space given by a modification of the lightcone constraint $C_X=\hat{X}^2 \sim R^2=\frac{\mathbf{C}_{2,j}}{m^2 \xi^{2/\omega}}$, with $\mathbf{C}_{2,j}$ denoting the quadratic Casimir operator which should have the following form:
\begin{equation}
    \hat{\mathbf{C}}_{2,j}=-\frac{1}{2}\xi^{2/\omega} \text{tr}(\hat{q}^2)
\end{equation}
Eigenvalue of the quadratic Casimir operator for arbitrary $\mathfrak{so}(1,d+1)$-algebra in embedded space is following:
\begin{equation}
    \mathbf{C}_{2,j}=\Delta_i(\Delta_i-D+2) \frac{(D-2)^2}{16}.
\end{equation}
where $\Delta_i$ is a canonical dimension of a scalar particle at the classical level. For example, in the $D=6$ case, one has $\mathbf{C}_{2,j}=-3$ for spinless classical particles, for $D=8$ it equals $-45/4$ and so on. 

The quantum analog of the $SO(1,d+1)$ charge can be represented as:
\begin{equation} \label{quantumQ}
    \hat{q}^{M N}_k= \frac{i}{\xi^{1/\omega}} \left(\hat{Z}^M_k, \partial_{Z,k}^N-\hat{Z}^N_k \partial_{Z,k}^M\right)
\end{equation}
here $\partial_{Z,i}^N$ is a partial derivative on $\hat{Z}_i$. 
The physical constraints $C_P$ at the quantum level can be defined as an equation on the eigenvalues of the Casimir operator which looks like the Klein-Gordon-Fock equation for the scalar field \cite{GromovAdS}:
\begin{equation} \label{Cas}
    \hat{\textbf{C}}_{2,i} ~ \widehat{\Psi}= \textbf{C}_{2,i} \widehat{\Psi}  
\end{equation}
The dual wave function $\widehat{\Psi}$ of the fishchain is given by the tensor product of free massive scalar on $AdS_{D-1}$ which is parameterised by $\hat{Z}^2_i=-1$. This equation is nothing  but the equation of motion for the wave function in the bulk of the corresponding AdS space. The part of the dual wave function that depends on the radius of AdS space-time is factorised and given by the Bessel function \cite{GromovAdS}.

Although expression (\ref{quantumQ}) is formally similar to the Hamiltonian in the classical form, it is not suitable for the quantized fishchain, since quantum $\hat{q}^{M N}_i$ are not irreducible symmetric tensors due to non-commutativity of the corresponding operators. This is why there is a need to specify a quantized form of the Hamiltonian. In this case, it is necessary to ''normal-order'' the charge operators of $SO(1,d+1)$ and to redefine  them as follows:
\begin{equation}\label{qno}
     :\hat{q}_{j}^{2}:^{N M}\;\equiv\;\left(\hat{q}_{j}^{2}\right)^{N M}-\frac{i\;d}{2\xi^{1/\omega}}\hat{q}_{j}^{N M}-\textbf{C}_{2,j}\frac{ d^{2}}{8D\xi^{2/{\omega}}}\eta^{N M} 
\end{equation}
This structure represents a traceless and symmetric tensor that satisfies the requirements imposed on the quantized charge density. 
Thus, the Hamiltonian at the quantum level can be written in the form:
\begin{equation}
    \hat{H}_q=\text{tr}\left(\prod_i \frac{:q^2_i:}{2}\right)-1
\end{equation}
This type of the ''normal-ordered'' operator provides integrability and correspondence between the quantum fishchain and the biscalar fishnet model, which will be seen in the following sections. By the action of such an operator on $\Tilde{\Psi}$:
\begin{equation} \label{Hqpsi}
    \hat{H}_q ~  \widehat{\Psi}(Z_i,\ldots,Z_j)=0
\end{equation}
one can unambiguously determine the spectrum and the dynamics of the chain of particles in this model. It is worth noting that the condition (\ref{Hqpsi}) imposed on the dual wave function violates unitarity: it is responsible for the appearance of tachyons in the spectrum, despite the fact that the Hamiltonian operator is self-adjoint.

\section{Test of $J=2$ spectra}

\subsection{6d fishnet model with deformation parameter $\omega=1$}

For $J=2$ the CFT wave function is fixed by conformal symmetry and represented in the following form \cite{GromovAdS}:
\begin{equation}
    \widehat{\Psi}_{\Delta,S}(Z_1,Z_2) =\frac{\left(\frac{Z_1 \mathcal{N}}{Z_1 \cdot \mathcal{X}}-\frac{Z_2 \mathcal{N}}{Z_2 \cdot \mathcal{X}}\right)^S}{(Z_1 \cdot \mathcal{X})^{\frac{(\Delta-S)}{2}} (Z_2 \cdot \mathcal{X})^{\frac{(\Delta-S)}{2}}} F\left(Z_1 \cdot Z_2, \log\frac{Z_1 \cdot \mathcal{X}}{Z_2 \cdot\mathcal{X}} \right)
\end{equation}
where $\mathcal{X}_{A}$ and $\mathcal{N}_{A}$ are the fixed null vectors in AdS$_{D-1}$ space  \cite{GromovAdS}. The whole design is very similar to the CFT wave function, and the denominator contains functions which can be called propagators near the boundary. As noted before, the dual CFT wave function must satisfy the constraints:
\begin{equation} \label{Cas}
    \hat{\textbf{C}}_{2,i} \circ  \widehat{\Psi}_{\Delta,S}(Z_1,Z_2)  = \textbf{C}_2  \widehat{\Psi}_{\Delta,S}(Z_1,Z_2), ~ \;i=1,2
\end{equation}
Two constraints (\ref{Cas}) can be rewritten in expanded form using their sum and difference, so after some tedious operations we have for an arbitrary dimension $d$:
\begin{equation}\label{firstc2}
    2 (\gamma+\cosh \kappa)F^{(1,1)}+(D-2-\Delta+S)F^{(0,1)}-(\Delta-S) \sinh \kappa F^{(1,0)}=0;
\end{equation}
\begin{multline}\label{secondc2}
    \left(\gamma ^2-1\right) F^{(2,0)}+F^{(0,2)}+F^{(1,0)} (\gamma  (D-\Delta +S-1)+\cosh \kappa (S-\Delta ))- \\ -2 \sinh (\kappa ) F^{(1,1)}+\frac{1}{4} F \left(\Delta  (-2 D+\Delta +4)+2 S (D-\Delta -2)+4  \textbf{C}_2 +S^2\right)=0
\end{multline}
where $F=F(\gamma, \kappa)$ so $\gamma=Z_1\cdot Z_2$ and $\kappa= \log\frac{Z_1 \cdot \mathcal{X}}{Z_2 \cdot\mathcal{X}} $.
These equations allow fixing the behaviour of the $F$ function in an unambiguous way near the boundary and the bulk of AdS$_{D-1}$ space. Unfortunately, they cannot be solved, although the behaviour near the boundary can still be described quite accurately. In the limit, where $\gamma \rightarrow \infty$ and $\kappa$ is fixed, one can see from the  equation (\ref{firstc2}) that $F$ should scale as $ \sim \frac{1}{\gamma}$. After the appropriate substitution into the second equation (\ref{secondc2}) one has:
\begin{equation}
    \left(\gamma\partial_\gamma+\vartheta_1\right)\left(\gamma \partial_\gamma + \vartheta_2\right)F(\gamma)=0
\end{equation}
where $\vartheta_{i}$ show how the solutions depend on the eigenvalue of the quadratic Casimir operators and $D$. For $d=6$ and $\omega=1$, one can obtain that
\begin{equation}
    \vartheta_1=-3-\Delta+S, ~\; \vartheta_2=15-\Delta+S.
\end{equation}
Inserting $F \sim \gamma^{-\vartheta_i}$ (in the bulk the regular one is the first) in the Hamiltonian after large $\gamma$, one gets:
\begin{equation}
    \left( \frac{(-\Delta +S+3) (-\Delta +S+5) (\Delta +S-1) (\Delta +S+1)}{16 \xi^4} \right) \widehat{\Psi}_{\Delta,S} =\widehat{\Psi}_{\Delta,S}
\end{equation}
The resulting spectrum is fully consistent with what we have obtained for the biscalar fishnet case in (\ref{spec61}). Thus, it can indeed be considered that the generalised fishchain model in AdS$_7$ is a dual representation of the biscalar fishnet model in six dimensions \footnote{ We have also checked that the relation for the spectra for $d=6, \omega=3/2$ isotropic fishnet model is coincide by putting the corresponding dual wave function in the fishchain Hamiltonian of the form: $H_q \sim \frac{1}{8} \text{tr}\left(:q_1 ^2:  ~ :q_1: \cdot :q_2:   ~ :q_2^2: \right)-1$.}.

\subsection{6d fishnet model with deformation parameter $\omega=2$}

.

Using equations (\ref{firstc2}) and (\ref{secondc2}) we can write the Hamiltonian for this interaction in the following form:
\begin{equation}
    \hat{H}_q=\text{tr}\left(\prod_{i=1}^J \left[ \frac{:q^2_i:}{2} \right]^2\right)-1
\end{equation} 
in the correspondence with the general procedure.
This Hamiltonian reproduces the desired spectrum exactly:
\begin{multline}
    \frac{1}{256 \xi^2}\left((\Delta -S-3) (\Delta -S-5) (\Delta -S-7) (\Delta -S-9) \right.  \\ \left. (\Delta +S-5)  (\Delta +S-3) (\Delta +S-1) (\Delta +S+1)\right) \widehat{\Psi}_{\Delta,S}=\widehat{\Psi}_{\Delta,S}
\end{multline}
Thus we confirm the correspondence between the biscalar fishnet in the six dimensions and the quantised fishchain model.

\section{The holographic map}

Although we have obtained spectra of the relevant models, a more consistent proof of the duality between fishchain and fishnet is still required. Since the fishnet models live on the boundary, there should be a limit of the fishchain wave function in the bulk. Introducing a fishchain wave function requires a transformation that must satisfy the equation of motion (\ref{Cas}) and conformal symmetry:
\begin{equation}
    \widehat{\Psi}_\mathcal{O}(\{Z_i\})=\int \prod_i \frac{d^d x_i}{-(2 \pi)^{d/2} (Z_i \cdot \mathcal{X}_i)^{\textbf{C}_2}} \Psi_\mathcal{O}(\{x_i\}),
\end{equation}
where $\mathcal{X}_i$ is the null vector whose components depend on $x_i$, the notation $\{x_i\}$  means a set of coordinates where $i=1 \ldots J$ and the lower index of the CFT wavefunction means that we consider it as an arbitrary correlator with the local operator $\mathcal{O}$ introduced in the first section. The eigenvalue of the Casimir operator is still denoted by $\textbf{C}_2$, it plays the role of mass in the ''equation of motion'', satisfying the local constraints imposed on the dual CFT wavefunction: $(Z_i \cdot \mathcal{X}_i)^{-\textbf{C}_2}$ can be thought of as a bulk-to-boundary propagator according to \cite{GromovAdS}.

The next task is to rewrite the corresponding integral in $SO(1,d+1)$ invariant terms. The integration measure can be defined in a covariant way generalising the results of \cite{Simmons-Duffin:2012juh}:
$$
D^{d+2}X=d^d X \delta(X^2),
$$
which corresponds to the measure of the projective lightcone on $\mathbb{R}^{1,d+1}$ \cite{GromovAdS}.  So the integral having the $SO(1,d+1)$ invariant measure can be written as: 
\begin{equation}
    \lim_{\Lambda \rightarrow \infty} \frac{1}{2 \Lambda}  \underset{-e^\Lambda < X^{+}< e^\Lambda}\int   \frac{d^d X \delta(X^2) }{ - (4 \pi)^{d/2}}\frac{f(X/X^{+})}{X^{+} (Z \cdot \mathcal{X})^{\mathbf{C}_2}} =\int D^d X\frac{ f(X/X^+) }{X^{+}(Z \cdot \mathcal{X})^{\mathbf{C}_2}}
\end{equation}
where $X^+$ is the lightcone coordinate, $\Lambda$ is the cut-off parameter.
Having such an uplifting procedure one can construct the fishchain wavefunction. It can be rewritten in the bulk as:
\begin{equation}\label{uppsi}
    \widehat{\Psi}_\mathcal{O}(\{Z_i\})=\int \prod_i \frac{D^{d+2} X_i}{-(2 \pi)^{d/2} (Z_i \cdot \mathcal{X}_i)^{\textbf{C}_2}} \psi_\mathcal{O}(\{X_i\})
\end{equation}
where the integrand is 
\begin{equation}
    \psi_\mathcal{O}(\{X_i\})=\frac{\Psi_\mathcal{O}(\{\vec{X}_i/X_i^{+}\})}{X^{+}_1 \ldots X^{+}_J} ,
\end{equation}
Thus, it becomes apparent that the dual wave function can be rewritten in terms of the coordinates of the space of dimension $D$.

The next step is to invert the graph-building operator with appropriate rules of differentiation (\ref{dif1}-\ref{dif2}):
\begin{equation}
    \hat{B}^{-1}=\prod_i |x_i - x_{i+1}|^{2\omega} \prod_j \frac{\left(-\square_j^{(d)}\right)^{\omega}}{2 \xi^{2}} 
\end{equation}
Again, we restrict ourselves to the integer or half-integer deformation parameter $\omega$ and omit the $\kappa$-factor because it is just an insignificant number. By means of properties of the integral (\ref{uppsi}) we can show that it is identical to the action of the quantized charge algebra of $SO(1,d+1)$.  Uplifting $d$-dimensional d'Alembertian to $D=d+2$: $\square^{(D)}=\square^{(d)}-\partial_{X_+}\partial_{X_-}$, one can show that
\begin{equation}
    \int \prod_i \frac{d^{d} x_i}{ (Z_i \cdot \mathcal{X}_i)^{\textbf{C}_2}} B^{-1} \circ \Psi(\{x_i\}) =\int \prod_i \frac{D^{d} X_i}{ (Z_i \cdot \mathcal{X}_i)^{\textbf{C}_2}} \mathbf{B}^{-1} \circ \psi(\{\vec{X}_i/X^{+}_i\})
\end{equation}
where the inverted graph-building operator acts on $X_i$ coordinates: 
\begin{equation}\label{invB}
    \mathbf{B}^{-1}=\prod_i (X_i \cdot X_{i+1})^\omega \prod_j \frac{\left(\square^{(D)}\right)^\omega}{2 \xi^{2}}
\end{equation}
From formula (\ref{invB}) it is obvious how the anomalous dimension $\Delta$ is scaled by the coupling constant. 
Then observing that the action of this operator $\mathbf{B}^{-1}$ corresponds to the action of the quantized charge density $\textbf{q}_i$  acting on $X_i$, one can conclude that
\begin{equation}\label{qBD}
    \text{tr}\prod_i \left( \frac{:\mathbf{q}_i^2:}{2}\right)^\omega \simeq \prod_i (X_i \cdot X_{i+1})^\omega \prod_j \frac{\left(-\square_j^{(D)}\right)^\omega}{2 \xi^{2}}= \mathbf{B}^{-1}
\end{equation}
i.e. these two operators can be identified because
\begin{equation}
    \mathbf{\hat{q}}_i^{MN}=\hat{X}^M_i \hat{K}^N_i - \hat{X}^M_i \hat{K}_i^N,
\end{equation}
where $K$ is the operator that acts on $X_i -$coordinates
$$\hat{K}_{M,i}=-\frac{i}{\xi^{\frac{1}{\omega}}} \frac{\partial}{\partial X^M_i},$$ so the ''normal ordered'' $:\textbf{q}_i^2:$ can be defined in the same way as (\ref{qno}) with the corresponding operators $X$ and $K$.
On the basis of equality (\ref{qBD}), it is clear that
\begin{equation}
\int\prod_{i=1}^{J}\frac{D^{d}X_{i}}{-(2\pi)^{d/2}(Z_{i}.X_{i})^{3}}\hat{\mathbf{q}}_i ~\psi_{\cal O}(\{X_i\})=\hat{q}_i\int\prod_{i=1}^{J}\frac{D^{d}X_{i}}{-(2\pi)^{d/2}(Z_{i}.X_{i})^{3}}\psi_{\cal O}(\{X_i\})
\end{equation}
where $q_i$ acts only on $Z_i$-coordinates; consequently, 
\begin{equation}
    \int\prod_{i=1}^{J}\frac{D^{d}X_{i}}{-(2\pi)^{d/2}(Z_{i}.X_{i})^{3}}\mathbf{B}^{-1}\psi_{\cal O}(\{X_i\})=\left(\hat{H}_q+1\right) \widehat{\Psi}(\{Z_i\}).
\end{equation}
This expression completes the proof of the correspondence. We can see that $B$ acting on wave function in $x_i$ coordinates is equivalent to the action of $q^{MN}_i$ acting on the bulk $Z_i$-coordinates. 
 
It remains only to construct explicitly the norm on the Hilbert space for the dual CFT wave function. As noted in the first section the CFT wave function can be defined as (\ref{cftwf}), so the off-shell norm of states (''length of vector'') can be expressed in the following form:
\begin{equation} 
    \langle\Psi_\mathcal{\Tilde{O}} | \Psi_\mathcal{O} \rangle= \int \prod_{i=1}^J \frac{d^d x_i}{(2\pi)^{d/2}}  \Psi^\dag_\mathcal{\Tilde{O}}(\{x_i\}) \prod_{j=1}^J (-\square_j)^\omega\Psi_\mathcal{O}(\{x_i\})
\end{equation}
In the uplifted coordinates, it can be rewritten as
\begin{equation}
    \langle\Psi_\mathcal{\Tilde{O}} | \Psi_\mathcal{O} \rangle= \int \prod_{i=1} \frac{D^d X_i}{(2 \pi)^{d/2}}   \Psi^\dag_\mathcal{\Tilde{O}}(\{X_i\}) \prod_{j=1}^J (-\square_j)^\omega\Psi_\mathcal{O}(\{X_i\}).
\end{equation}
So after such a redefinition we can find norm for a fishchain wave function
\begin{equation}
    \langle\widehat\Psi_\mathcal{\Tilde{O}} | \widehat\Psi_\mathcal{O} \rangle = \int_{Z_i^+ < \Lambda_i} \prod_{i=1}^J \frac{d^{d+1}Z_i}{(2\pi)^{d/2}} \widehat\Psi^\dag_\mathcal{\Tilde{O}} (\{Z_i\}) \prod_j \left(\nabla^\dag_{j,M}  \nabla_{j}^M + \textbf{C}_{2,j}\right)^\omega \widehat\Psi_\mathcal{\Tilde{O}}(\{Z_j\})
\end{equation}
with the integration measure $d^{d+1}Z_i=d^{d+2} Z_i \delta(Z_i^2-1)$ on unit AdS radius and the differential operator
\begin{equation}
    \nabla_{i,M}=\frac{\partial}{\partial Z_{i}^M}+ Z_{i,M} \left(Z_i \cdot \frac{\partial}{\partial Z_{i}}\right)
\end{equation}
which is just the covariant derivative on AdS$_{d+1}$.  Thus, the behaviour of the norm of the wave functions in the AdS bulk is exactly the same as in \cite{GromovAdS}.
It is also possible to construct similar Hilbert spaces for the correlation functions in full agreement with the wave functions described above for the triangular fishnet lattice.

\section{Conclusion}

In this article, we have generalized the results of the papers \cite{GromovSever,GromovAdS} on the fishchain to the cases of higher dimensions and general interaction. We also have found that the shape of the scaling dimension at the classical level for the correlator in the fishchain model also allows us to draw parallels with the t'Hooft holographic principle. At the quantum level, we have seen the same thing: here also emerges a fixed radius of the anti-de-Sitter space during quantization of the fishchain proportional to eigenvalue of the quadratic Casimir operator. The duality is proved quite easily: this is based entirely on properties of the graph-building operator and the quantized charge density of the $SO(1,d+1)$ group. 

We also managed to obtain a spectrum of the non-isotropic fishnet lattice for the 6d non-isotropic biscalar fishnet and fishchain models. Surprisingly, the dual model passes the tests quite well reproducing all spectra. We have seen that the fishchain model is a non-unitary theory, and tachyonic states are detectable in it from the appearance of the resulting spectra on the quantum level due to an analog of the level-matching condition in string theory arising from the constraint (\ref{hamiltclas}).  

It would be interesting to see how a six-dimensional theories with a hexagonal and triangular structure of Feynman diagrams in the planar limit, introduced by Zamolodchikov \cite{Zamolodchikov}, might fit within these models.

There are also many possibilities  which way to go with multi-dimensional fishchain models. Let us list them in more detail.

There is a fairly strong assumption that in the continuous limit $J\rightarrow \infty$  this kind of model \cite{Bassosigma,Alfimov} should  be modified into a non-linear sigma model on AdS. The equations of the fishnet and sigma models are superficially similar, and the TBA \cite{Bassosigma} for the four-dimensional fishnet model has the similar form as for the AdS sigma model in the thermodynamic limit, however, proving that the ads sigma model is a fishchain model in the continuous limit seems like an interesting task. Nevertheless, such a proof would open the way to direct quantization of sigma models and the study of complicated string models \cite{Bajnok:2008it}.

Also, it is a curious task to consider fishchain models that include fermionic fields (such a fishnet models are widely represented and extensively studied in the literature, e.g. \cite{Kazakov:2018gcy,Pittelli:2019ceq}), or, moreover, to study fishchain models that have supersymmetry in the bulk.

It seems extremely fruitful to consider SYK models as they look similar to the one-dimensional biscalar fishnet model \cite{Polchinski2016} and derive dual fishchain counterparts, as the AdS$_2$/CFT$_1$ correspondence is well researched \cite{Polchinski2016}. It should also be noted that for $d=1$ and $\omega=1/4$ in (\ref{spectrumforall}) the fishnet model has a very similar spectrum. It is possible to find the corresponding fishchain model and compare it with the SYK-fishchain as they seem to be close to each other due to isomorphism between the $SO(2,1)$ and $PSL(2,R)$ groups. There are also low-dimensional models that are defined by spin chain models for example Lipatov's reggeized gluon theory in the leading order, for which dual models can also be obtained \cite{Alfimov}. One might expect their dual description to be more complicated, but also, more practical and curious.

\subsubsection*{Acknowledgements}
We would like to thank Andrey Onishchenko for many valuable comments and discussions. Also, we are grateful to Grigory Korchemsky, Fyodor Levkovich-Masliuk and Nikolay Gromov for their clarifications. Financial support from the Russian Science Foundation grant no. 21-12-00129 is cordially acknowledged.

\subsection*{A. Spectrum of non-isotropic lattice for biscalar fishnet theory}

The star-triangle identity \cite{Chicherin} is given by

\begin{equation}
    \int d^d x_0 ~ x_{01}^a x_{02}^b x_{03}^c=\pi^{d/2} \frac{\Gamma \left(\frac{a}{2}+\frac{d}{2}\right) \Gamma \left(\frac{b}{2}+\frac{d}{2}\right) \Gamma \left(\frac{c}{2}+\frac{d}{2}\right) }{\Gamma \left(-\frac{a}{2}\right)\Gamma \left(-\frac{b}{2}\right) \Gamma \left(-\frac{c}{2}\right)} x^{-a-d}_{12} x^{-b-d}_{23} x^{-c-d}_{13}
\end{equation}
and when $x_3 \rightarrow \infty$ one can simplify it up to
\begin{equation}
    \int d^d x_0 ~ x_{01}^a x_{02}^b = \pi^{d/2} \frac{\Gamma \left(\frac{a}{2}+\frac{d}{2}\right) \Gamma \left(\frac{b}{2}+\frac{d}{2}\right) \Gamma \left(-\frac{a}{2}-\frac{b}{2}-\frac{d}{2}\right)}{\Gamma \left(-\frac{a}{2}\right) \Gamma \left(-\frac{b}{2}\right) \Gamma \left(\frac{a}{2}+\frac{b}{2}+d\right)} x_{12}^{a+b+d}
\end{equation}
Then we can use this identity in this form to obtain the spectrum of the wave function for $J=2$. Conformal symmetry fixes the form of the CFT wave function as follows
\begin{equation}
    \Psi_{S,\Delta,x_0}(x_1,x_2)=\frac{x_{12}^{\Delta-S-d+2\omega}}{x_{01}^{\frac{\Delta-S}{2}} x_{02}^{\frac{\Delta-S}{2}}} \left(2\frac{(n x_{02})}{x_{02}^2}-2\frac{(n x_{01})}{x_{01}^2}\right)^S
\end{equation}
This wave function plays the role of the eigenvector of the graph-building operator. So the eigenvalues of the operator can be defined as follows
\begin{equation}
    \mathcal{H} \circ \Psi_{S,\Delta,x_0}=h_{\Delta,S}^{-1}\Psi_{S,\Delta,x_0}
\end{equation}
and the graph-building operator is given in a generalised form 
\begin{equation}
     \mathcal{H}=\int d^dx_1 d^dx_2 ~ \frac{1}{x_{12}^{4\omega} x_{13}^{d-2\omega} x_{24}^{d-2\omega}}
\end{equation}
Using inversion around $x_0$, the star-triangle relation and integration by parts as it was done in \cite{ ExactCorr, FishnetAnyDim}, a spectrum for the non-isotropic lattice integral can be obtained.

\begin{equation}
  h_{\Delta,S}= \frac{   \Gamma (\omega )^2}{\Gamma \left(\frac{d}{2}-\omega \right)^2 } \frac{ \Gamma \left( \frac{d}{4}+\frac{S}{2}-\frac{\Delta}{2} \right) \Gamma \left(\frac{d}{4}  -2 \omega  + \frac{S}{2} + \frac{\Delta}{2}\right)}{\Gamma \left(\frac{d}{4}+\frac{S}{2}+\frac{\Delta}{2}\right) \Gamma \left(\frac{d}{4}  +2 \omega  + \frac{S}{2} - \frac{\Delta}{2} \right)}
\end{equation}
In the $\omega=d/4$ limit, the result of \cite{FishnetAnyDim} can be restored. 
\bibliographystyle{unsrt}
\bibliography{refs}
\end{document}